\documentstyle[aps,epsfig]{revtex}
\draft
\begin{document}
\def\la{\langle}
\def\ra{\rangle}
\def\om{\omega}
\def\Om{\Omega}
\def\vep{\varepsilon}
\def\wh{\widehat}
\newcommand{\beq}{\begin{equation}}
\newcommand{\eeq}{\end{equation}}
\newcommand{\beqa}{\begin{eqnarray}}
\newcommand{\eeqa}{\end{eqnarray}}
\newcommand{\intf}{\int_{-\infty}^\infty}
\newcommand{\into}{\int_0^\infty}
%

%      Title page and abstract...
\begin{title}
{\Large\bf Sources of quantum waves}
\end{title}
\author{\large A. D. Baute$^{1,2}$, I. L. Egusquiza$^1$ and J. G. Muga$^2$}
\address{
${}^1$ Fisika Teorikoaren Saila,
Euskal Herriko Unibertsitatea,
644 P.K., 48080 Bilbao, Spain\\
${}^2$ Departamento de Qu\'\i mica-F\'\i sica,
Universidad del Pa\'\i s Vasco, Apdo. 644, Bilbao, Spain
}
\maketitle

\begin{abstract}
Due to the space and time dependence of the wave
function in the time dependent Schr\"odinger equation,
different boundary conditions
are possible. The equation is usually  solved
as an ``initial value
problem'', by fixing the value of the wave function in
all space at a given instant. We compare this standard approach to
``source boundary conditions'' that
fix the wave at all times in a given region, in particular at
a point in one dimension.
In contrast to the well-known physical interpretation of the
initial-value-problem approach,
the interpretation of the source approach has remained unclear,
since it introduces negative energy
components, even for ``free motion'', and a time-dependent norm.
This work provides physical meaning to the source
method by finding the link with equivalent initial value problems.
\end{abstract}

\pacs{PACS: 03.65.-w
\hfill  EHU-FT/0007}

\section{Introduction}
The time dependent Schr\"odinger equation is commonly solved as
an ``initial value problem''.
In one dimension this means that the wave function $\psi(x,t)$
is fixed for all $x$ at a reference time, usually $t=0$.
However, it is also possible to use different boundary conditions
by fixing $\psi(x,t)$ at a space point, say $x=0$, at all
times. This  ``source'' approach has been followed
to study the arrival time in quantum mechanics \cite{Allcock:1969a}, or
characteristic propagation velocities and times (``tunneling times'')
in evanescent conditions
\cite{Stevens:1980,Stevens:1983,Moretti:1992,Ranfagni:1991,Buettiker:1998a,%
Buettiker:1998b,Muga:2000}.
In all these applications the wave function is assumed to vanish for
$x\ge 0$ and $t<0$,
\beq\label{3.1}
\psi_s(x,t)=0,\;\; x\ge 0,\;\; t<0,
\eeq
the potential is zero or constant for $x\geq0$ and $t>0$, and the solution
of the Schr\"odinger equation is sought for the positive half-line
and positive times, $\psi_s(x\ge 0,t>0)$.
We shall also limit the present discussion to these
premises, and use the subscript $s$
to denote the solutions that obey  Eq.~(\ref{3.1}),
computed for the space-time domain
$D_+\equiv\{x\ge0;t>0\}$.
They have been invariably
written, up to constant factors
and notational changes, as
\beq\label{source}
\psi_s(x,t)=\frac{1}{h^{1/2}}\intf dE\, e^{ixp_+/\hbar-iEt/\hbar}
\chi_s(E),
\eeq
where
\beq\label{eft}
\chi_s(E)=\frac{1}{h^{1/2}}\intf dt\, \psi_s(x=0,t) e^{iEt/\hbar}
\eeq
is the energy Fourier transform of the given ``source signal''
$\psi_s(x=0,t)$ and
$p_+=(2mE)^{1/2}$,
with the branch cut taken along the negative imaginary axis of $E$.
Because of Eq.~(\ref{3.1}), $\chi_s(E)$ is an analytical
function of $E$ for ${\rm Im} E>0$; however, this does not guarantee the
absence of singularities of $\chi_s(E)$ for real $E$, additional
to the branch point. In fact, the integral over $E$ is to be understood
as going above the real axis; alternatively, $E\to E+i0$ is to be
substituted in the expressions above.

It is easy to check by substitution that  Eq.~(\ref{source})
satisfies the Schr\"odinger equation.
Actually Allcock derived Eq.~(\ref{source}) as the most general form compatible
with  Eq.~(\ref{3.1}) \cite{Allcock:1969a}.
However he did not analyze its connection with
initial value problems, or the meaning of the negative
energies, so its physical interpretation has remained unclear.
What are the negative energy components? How should
such a wave be normalized? In time instead of position? Indeed, should
it be normalized? Is the source
to be interpreted as a multi-particle source?
Except for the detailed analysis by Allcock, most authors have used
the source boundary conditions without discussion, or as a simplifying
approach whose actual physical content was questionable, as reflected,
for example, in Moretti's words \cite{Moretti:1992}:
``{\it The model is surely more suited in the electromagnetic
case; for a quantum particle, it seems more satisfactory a localization in a
given region, at a given time, rather than a ``creation'' during a given time
interval, at a fixed position. [....] {\rm (The model)} 
is chosen by reasons of simpler
mathematics}''.

A ``source signal'' such as
\cite{Stevens:1980,Stevens:1983,Moretti:1992,Ranfagni:1991,Buettiker:1998a,%
Buettiker:1998b,Muga:2000}
\beq\label{sig}
\psi_s(x=0,t)=e^{-i\om_0 t}\Theta(t)
\eeq
leads to an infinite norm as $t\to\infty$ for $\om_0>0$. An interpretation
of such an infinite norm might only be possible by invoking a
multi-particle approach which we shall not pursue here. Instead, if we
keep within the one particle picture, it becomes clear that not
all ``signals'' $\psi_s(x=0,t)$ may be allowed physically.
This does not mean that the results obtained with  Eq.~(\ref{sig}) are
meaningless. This signal may be an idealized approximation
during a certain time interval of an actual signal, and 
it is also an elementary component of an arbitrary physical signal.
Our basic philosophy here is that the physical interpretation
of  Eq.~(\ref{source}) must rest on a process where a wave packet
located in the left half-line is liberated at $t=0$ and
crosses, in general only partially, to the right half-line.
This was the point of view of Allcock too \cite{Allcock:1969a},
although our scope will differ from his work. There must be a link
between a physically valid $\psi_s$ and
some initial value problem, and the objective of this paper is to find
and discuss such a link and the meaning of the
negative energies in  Eq.~(\ref{source}). The consequences on the possibility to
define
time-of-arrival distributions in quantum mechanics are also examined.

\section{Equivalence between initial-value and source boundary
conditions}
\subsection{Free motion on the full line}

We shall solve in this section the initial value
problem corresponding to a wave restricted fully to the left half-line
up to $t=0$, in the interval $[a,b]$, with $a<b<0$,
by an appropriate time-dependent potential that vanishes
for $t\ge 0$.
The resulting wave $\psi_{iv}$ is only free from $t=0$ onwards.
It is also useful to introduce
a related wave $\psi_f$ which evolves freely {\it at all times},
positive and negative,
and is equal to $\psi_{iv}$ for $t\ge 0$,
\beq
\psi_f(x,t)=\int_a^b dx' \la x|U(t,t'=0)|x'\ra \psi_{iv}(x',t'=0), \qquad
\forall t\,.
\eeq
Here
$U(t,t')=\exp[-iH_0 (t-t')/\hbar]$ is the unitary evolution operator
for the free-motion Hamiltonian $H_0$,
\beq
\la x|U(t,t')|x'\ra=
\left[\frac{m}{ih(t-t')}\right]^{1/2}e^{im(x-x')^2/2\hbar (t-t')}.
\eeq
By setting the following source signal,
\beq\label{ic}
\psi_s(x=0,t)=\psi_{iv}(x=0,t)\Theta(t)=\psi_f(x=0,t)\Theta(t),
\eeq
the corresponding $\psi_s(x,t)$ is obtained via  Eq.~(\ref{source}).
We shall now establish the equality between $\psi_s$ and $\psi_f$
for positive times and positions when the ``boundary condition'' of  
Eq.(\ref{ic}) is imposed on $\psi_s$.
$\psi_f(x,t)$ may also be written as a momentum integral,
\beq\label{8}
\psi_f(x,t)=\frac{1}{h^{1/2}}\intf dp\,  e^{ixp/\hbar}
e^{-iEt/\hbar}\tilde\psi_f(p),
\eeq
where $E=p^2/2m$, and 
\beq\label{psip}
\tilde\psi_f(p)=\frac{1}{h^{1/2}}\int_a^b dx\,e^{-ipx/\hbar}
\psi_f(x,t=0).
\eeq
It can be separated into positive and negative momentum
components,
$\psi_f(x,t)=\psi_{f,+}(x,t)+\psi_{f,-}(x,t)$, where
\beq
\psi_{f,\pm}(x,t)=\frac{1}{h^{1/2}}\intf dp\, e^{ixp/\hbar}
e^{-iEt/\hbar}\tilde\psi_f(p) \Theta(\pm p).
\eeq
The positive momentum integral is now rewritten
in terms of the energy variable $E=p^2/2m$,
\beq\label{psi+}
\psi_{f,+}(x,t)=\frac{1}{h^{1/2}}
\into dE\,e^{ipx/\hbar}e^{-iEt/\hbar}\left(\frac{m}{2E}\right)^{1/2}
\tilde\psi_f(\sqrt{2mE})\,.
\eeq
To put $\psi_{f,-}$ in a similar form,  note that, because of the restriction
of the initial state within
the left half-line at $t=0$, its momentum
representation $\tilde\psi_f(p)$, Eq.~(\ref{psip}), 
is analytical in the complex plane $p$,
and decays to zero as $|p|\to\infty$ in the upper half-plane.
These two properties and the decaying behavior
of the exponentials of  Eq.~(\ref{8}) in the second
quadrant allow to substitute the integral along the negative real axis
by an integral along the positive imaginary axis (there are no poles and
the integral along a
large arc at infinity is zero),
\beq
\psi_{f,-}(x,t)=\frac{i}{h^{1/2}}\int_\infty^0   d\gamma e^{-x\gamma/\hbar}
e^{i\gamma^2t/2m\hbar}\tilde\psi_{f}(i\gamma)\;\;\;\;\; x\ge 0, t>0,
\eeq
where we have used $p=i\gamma$ $(\gamma\ge 0)$, and
$\tilde\psi_{f}(i\gamma)$ is {\it defined} through 
Eq.~(\ref{psip}). 
Changing now to the energy variable $E=-\gamma^2/(2m)$,
\beq\label{psi-}
\psi_{f,-}(x,t)=-\frac{i}{h^{1/2}}\int_{-\infty}^0 dE\,
e^{ixp_+/\hbar}e^{-iEt/\hbar} \left(\frac{m}{-2E}\right)^{1/2}
\tilde\psi_{f}\left[i(-2mE)^{1/2}\right]
\,,\;\;\;\;\;x\ge 0, t>0.
\eeq
Adding  Eq.~(\ref{psi+}) and Eq.~(\ref{psi-}) we can now write
$\psi_f(x,t)$ for $x\ge 0$, $t>0$ in the form of the right hand side of
Eq.~(\ref{source}),
\beq\label{14}
\psi_f(x,t)=\frac{1}{h^{1/2}}\intf dE\,e^{ixp_+/\hbar}e^{-iEt/\hbar}\chi_f(E),\;\;\;\;
x\ge 0, t>0,
\eeq
where
\beq\label{psife}
\chi_f(E)=(m/2E)^{1/2} \tilde\psi_{f}\left[(2mE)^{1/2}\right].
\eeq
As before, the cut defining the square root branch is taken at the negative
imaginary axis of $E$, so that the square roots are positive imaginary
numbers for negative energies.
It is also useful to write Eq.~(\ref{14}) as an integral in the momentum plane,
\beq\label{psipl}
\psi_f(x,t)=\frac{1}{h^{1/2}}\int_L dp\,  e^{ipx/\hbar} e^{-iEt/\hbar}
\tilde\psi_f(p)\,,\;\;\;\; x\ge 0, t>0,
\eeq
where the contour is an $L$-shaped line that goes downwards along the
positive imaginary axis and rightwards along the positive real axis.
Let us now
{\it define} the function $\phi$ by the integral
in Eq.~(\ref{psipl}) without any restriction on $x$ or $t$,
\beq\label{fi}
\phi(x,t)\equiv\frac{1}{h^{1/2}}\int_L dp\,  e^{ipx/\hbar} e^{-iEt/\hbar}
\tilde\psi_f(p),
\eeq
and analyze its behavior for $x\ge 0$ and $t<0$.
The contour $L$ can be closed
by a large arc in the first quadrant which does not contribute to the
integral. By Cauchy's theorem, $\phi(x,t)$ is zero for $x\ge 0$ and
$t<0$. In particular, this means that $\phi(x=0,t)=\psi_s(x=0,t)$.

We may also write Eq.~(\ref{fi}) as an integral along the real energy
axis,
\beq\label{psie}
{\phi}(x,t)=\frac{1}{h^{1/2}}\intf dE\, e^{ixp_+/\hbar}e^{-iEt/\hbar}
\chi_f(E),
\eeq
and solve for $\chi_f(E)$ by Fourier transform,
\beq
\chi_f(E)=\frac{1}{h^{1/2}}\intf dt e^{iEt/\hbar} \phi(x=0,t).
\eeq
However $\phi(x=0,t)=\psi_s(x=0,t)$, so that $\chi_f(E)=\chi_s(E)$,
and therefore $\psi_s(x,t)=\psi_f(x,t)$ in the space-time domain $D_+$
when the boundary condition specified in Eq.~(\ref{ic}) is satisfied
at $x=0$, and the state is initially localized as indicated above.

Since in many instances the given function is the signal $\psi_s(x=0,t)$
rather than its Fourier transform, it is also convenient to express
Eq.~(\ref{source}) as an integral over time in terms of the signal.
By inserting
Eq.~(\ref{eft}) and changing the order of integration, we find the
appealing expression
\beq\label{psisK}
\psi_s(x,t)=\int_0^t dt' K_+(t,x;t',x'=0) \psi_s(x'=0, t'),
\eeq
where the kernel is given by
\beq
\label{K+}
K_+(t,x; t',0)=\frac{1}{h}\intf dE\, e^{ip_+x/\hbar} e^{iE(t'-t)/\hbar}
=\frac{1}{hm}\int_L dp\, p\, e^{-ip^2(t-t')/(2m\hbar)}
e^{ipx/\hbar}.
\eeq
Notice that $K_+(t,0; t',0)=\delta(t'-t)$. Let us evaluate
Eq.~(\ref{K+}) for $x\ge 0$ along the $L$-shaped contour
in the momentum plane. For $(t-t')<0$ the
contour may be closed
by a large arc at $\infty$ in the first quadrant, which does not
contribute to the integral. Since no pole is
enclosed, the integral
is zero. This explains the upper limit $t$ in Eq.~(\ref{psisK}).

For $(t-t')>0$ the part of the contour in the imaginary axis
may be deformed into the negative real axis. This allows the
identification of $K_{+}$ with the derivative of the ordinary propagator,
\beqa\label{K2}
K_+(t,x;t',0)&=&
\intf dp\, \frac{p}{hm} e^{ip^2(t'-t)/(2m\hbar)}
e^{ipx/\hbar}
\\
\nonumber
&=&\frac{\hbar}{mi}\frac{d}{dx}\la x|U(t,t')|0\ra
\\
\nonumber
&=&\left[\frac{m}{ih(t-t')^3}\right]^{1/2}xe^{\frac{imx^2}{2\hbar(t-t')}}
\;\;\;\;
\;\;\;\; ({t>t',x\ge 0}).
\eeqa
A related result was obtained by Allcock on his way to
Eq.~(\ref{source}) \cite{Allcock:1969a}, by considering the retarded free
propagator. The expression of Allcock associates the source wave function
to a time integral of the wave function {\it at} the source point
times the free propagator, with a double-sided spatial derivative in
between. It can be shown that it is equivalent to Eq.~(\ref{psisK}). An
analogous relation is also obtained in three dimensions,
with similar techniques but
rather different objectives, by Mangin-Brinet et al
\cite{Mangin-Brinet:1998ck},
relating the wave function in one region of spacetime with the
convolution over that region of the initial wave function with the
propagator, plus a surface term with a double-sided gradient. 

The relation found between $\psi_s$ and $\psi_f$, and
Eq.~(\ref{psife}) are the main results of this
section. They provide a physical interpretation of the source
wave function in terms of an initial value problem when there is 
free motion after the initial state is released at 
$t=0$.
Note that the knowledge of the signal $\psi_s(x=0,t)$,
or of its Fourier transform
$\chi_s(E)$, allows to recover the negative momentum part of the
initial state $\psi_{iv}(t=0)$ for the corresponding initial value problem
only insofar as the amplitude $\tilde\psi_f(p)$, known in principle
along the positive real and imaginary axis, can be analytically
continued into the negative real axis.

In this way, the normalizability required of $\psi_{f}$ (or
$\psi_{iv}$, for that matter), is translated into a property of
$\psi_{s}$, which thus acquires the usual meaning of a wave-function
in quantum mechanics, that is, as a probability amplitude. It should
be observed that not all functions of the form of Eq.~(\ref{source}) can have
this meaning: only those that are indeed related to a given
$\psi_{iv}$ or $\psi_{f}$ can be thus understood.

We may now see the origin of the negative energies
in $\psi_s$: the contribution of the evanescent waves
in the source solution, Eq.~(\ref{source}),  is exactly equal to
the contribution of negative momenta in the
corresponding freely moving wave packet $\psi_f$, Eq.~(\ref{8}).
The change of variable and the shift in the path of integration lead
to what could be termed apparent evanescent waves, which in fact can
be understood as an artifact of the Fourier-Laplace transformation in
time.
It may be surprising from a classical perspective that such a
negative momentum contribution
exists at positive times and positions, considering that the wave packet is
entirely
localized on the left at $t=0$. In quantum mechanics, however,
the negative momentum (equivalently, evanescent or negative energy)
contribution is always present, since $\tilde\psi_{f,-}(p)$ is restricted
to a half line in momentum space and therefore $\psi_{f,-}(x)$
is necessarily different from zero from $-\infty$ to $+\infty$
except possibly at a finite number of points. It is true however
that the influence of
$\psi_{f,-}$ at $x>0$ diminishes with the distance to the origin.
This is illustrated in Figures 1 and 2, that show the probability
densities $|\psi_f(x,t)|^2$ and $|\psi_{f,\pm}(x,t)|^2$ versus $x$ at
two different instants. The state is initially (at $t=0$)
the ground state of an infinite well between $a$ and $b$,
$a<b<0$, and has zero average momentum, $\la p\ra=0$.
Its time evolution may be expressed in terms of know
functions, see the Appendix.
Note the long tails of $|\psi_{f,\pm}(x,t=0)|^2$
beyond $[a,b]$ in Figure 1, and the important interference pattern
between $\psi_{f,+}$ and $\psi_{f,-}$ in Figure 2.

%%%%%%%%%%%%%%%%%%%
\begin{figure}[h]
    \epsfysize=8cm 
    \centerline{\epsfbox{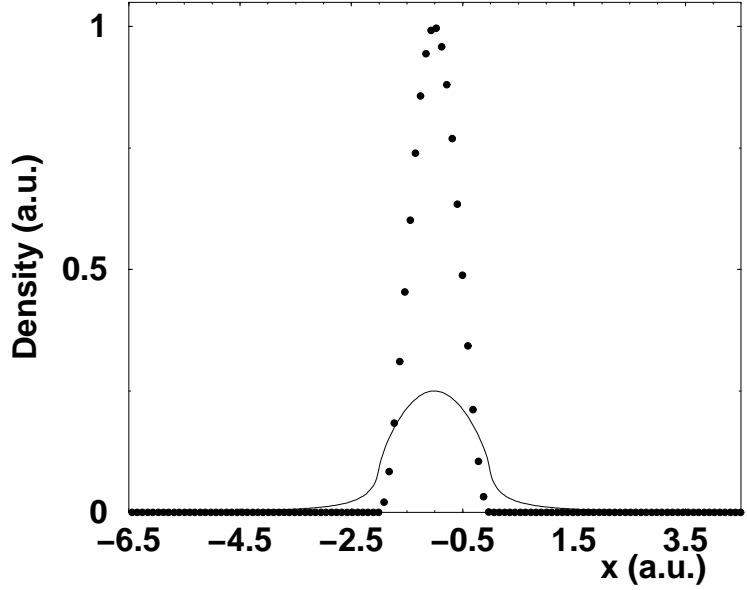}}
    \caption{\small $|\psi_f(x,t)|^2$ (dotted line), $|\psi_{f,+}(x,t)|^2$ (solid line), and 
$|\psi_{f,-}(x,t)|^2$ (dashed line, indistinguishable from the solid line)
versus $x$ for
$t=0$. The initial state is the ground state of an infinite well  
between $a=-2.01$ and $b=-0.01$, see  Eq.~(\ref{psiiv}). Here and in all 
figures, atomic units (a.u.) are used, $m=1$, $\hbar=1$.}
\end{figure}
%%%%%%%%%%%%%%%%%%%%
\begin{figure}[h]
 \epsfysize=8cm 
 \centerline{\epsfbox{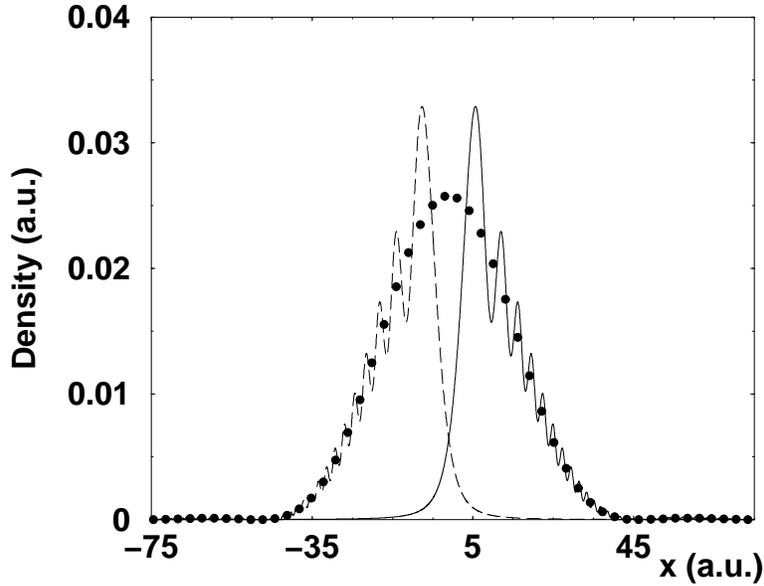}}
 \caption{\small Same as Fig. 1, but for $t=10$.}
\end{figure}
%%%%%%%%%%%%%%%%%%%%

\subsection{Generalization for potentials that vanish outside an interval}

The above results can be  generalized for an arbitrary
cut-off potential profile that vanishes outside an interval $[c,d]$,
located between the initial state
and the source point $x=0$, i.e. such
that  $a<b\le c\le d\le 0$.
The time evolution of the wave, $\psi_b$, for $x\ge d$ is given by
\cite{Brouard:1996,Baute2000}
\beq\label{psib}
\psi_b(x,t)=\frac{1}{h^{1/2}}
\int_\Omega dp\,e^{ipx/\hbar}e^{-iEt/\hbar}T(p)\tilde\psi_b(p)\;\;\;\;\;
x\ge d,
\eeq
(to be compared with Eq.~(\ref{8})),
where $T(p)$ is, for $p>0$,  the transmission amplitude, and
its analytical continuation elsewhere; similarly to 
Eq.~(\ref{psip}), $\tilde\psi_b(p)=h^{-1/2}\int_a^b dx\,e^{-ipx/\hbar}\psi_b(x,t=0)$.
The contour $\Omega$ goes from $-\infty$ to $\infty$. If there are bound states
it passes above the
corresponding poles of $T(p)$ in the positive imaginary axis.
$T(p)$ is analytical
in the upper half plane except in these points.
Moreover it tends to one
as $|p|\to\infty$ \cite{relax,Faddeev:1964,Newton:1980kh,Marinov:1996kb}.
This means that the same manipulations done for the
free motion wave function to write it in the form of $\psi_s$
can now be performed, but
instead of $\tilde\psi\left[(2mE)^{1/2}\right]$ in Eq.~(\ref{psife}) one must 
write $T[(2mE)^{1/2}]\tilde\psi\left[(2mE)^{1/2}\right]$ in the more
general case. The same arguments that
have enabled us to identify $\psi_f$ and $\psi_s$ in $D_+$
when the boundary condition of Eq.~(\ref{ic})
is satisfied are now applicable to identify
$\psi_s(x,t)$ and $\psi_{b}(x,t)$ in the same domain
when
\beq
\psi_s(x=0,t)=\psi_{b}(x=0,t)\Theta(t)
\eeq
is imposed as the signal at the source point $x=0$.

In any case free motion or motion with a localized potential barrier
in the full line do not easily allow to 
construct source signals dominated by
negative energy components.
But these are precisely the signals of interest
to study propagation velocities and characteristic
times for evanescent conditions
\cite{Stevens:1980,Stevens:1983,Moretti:1992,Ranfagni:1991,Buettiker:1998a,%
Buettiker:1998b,Muga:2000}.
The way to construct them
and their relation to initial value problems are examined in the next
subsection.

\subsection{``Step potentials'' with different asymptotic levels}
Here we shall show the connection
between source wave functions
and initial value problems for
the simple step potential,
\beq\label{stepp}
V(x)=\Theta(x) V_0,\;\;\;\;\;V_0>0, 
\eeq
as well as any other cut-off ``step'' potentials between $c$ and $d$,
such that $V(x>d)=V_0$ and $V(x<c)=0$.
As before, we assume that the initial state is localized between
$a$ and $b$, with $a<b\le c \le d \le 0$.
Following the method used in \cite{Brouard:1996} to write Eq.~(\ref{psib}),
it is possible to write the
wave function for $x>0$, after lengthy manipulations \cite{Baute2000},  as
\beq\label{exact}
\psi_{step}(x,t)=\frac{1}{h^{1/2}}\int_\Omega dp\, e^{-iEt/\hbar}e^{iqx/\hbar}
T^l(p)\tilde\psi_{step}(p)\,,
\eeq
where $\tilde{\psi}_{step}(p)=h^{-1/2}\int_a^b dx e^{-ipx/\hbar}\psi_{step}(x,t=0)$. 
This is an exact expression, where
$p$ is the momentum with respect to the lower level of the step;
$E=p^2/2m$, and
$q$ is the momentum with respect to the upper level. For real $p$, it is
given by
\beq
q=\cases{
-(p^2-p_0^2)^{1/2}&$\,\,\, p<-p_0$,\cr
i(p_0^2-p^2)^{1/2}&$\,\,\, -p_0<p<p_0$,\cr
(p^2-p_0^2)^{1/2}&$\,\,\, p>p_0$.\cr}
\eeq
Here and in the following $p_0$ equals $(2mV_0)^{1/2}$.
For arbitrary values of $p$ in the complex plane, $q=(p^2-p_0^2)^{1/2}$
with the understanding that the
two branch points at $p=\pm p_0$ are joined by the shortest branch cut, slightly
displaced below the real
axis. In particular, for a point $p=i\gamma$ ($\gamma>0$)
the corresponding $q$ is  $i(\gamma^2+p_0^2)^{1/2}$.
$T^l(p)$ is, for $p>p_0$,
the transmission amplitude for left incidence,
and its analytical continuation elsewhere.
As in the previous subsection, the contour $\Omega$ goes from $-\infty$ to
$\infty$ passing above the branch cut and possible bound-state poles of
$T^l(p)$ on the imaginary axis.

The integral in Eq.~(\ref{exact}) may also be written in the $q$-plane as
\beq
\psi_{step}(x,t)=\frac{1}{h^{1/2}}\int_C dq\, e^{iqx/\hbar}e^{-iEt/\hbar}
\frac{q}{p}T^l(p)\tilde\psi_{step}(p),
\eeq
where now $p=(q^2+p_0^2)^{1/2}$ has two branch points at
$q=\pm ip_0$ joined by a branch cut. The  contour $C$
goes from $-\infty$ to $+\infty$ passing above the
part of branch cut in the positive imaginary axis and above the
poles of $T^l(p)$ (Note that whenever there are bound-state poles they are
above the
branch point at $q=ip_0$).
It is possible, as in the previous section,
to deform the contour in the second quadrant
so that the final contour has an $L$ shape, going from $i\infty$ to zero
passing to the right of the poles, to the right of the branch cut
($p>0$ in the interval from $q=ip_0$ to
$q=0$), and from zero to
$\infty$ along the positive real axis, see Figure 3, 
\beq
\psi_{step}(x,t)=\frac{1}{h^{1/2}}\int_L dq\, e^{iqx/\hbar}e^{-iEt/\hbar}
\frac{q}{p}T^l(p)\tilde\psi_{step}(p).
\eeq
%
%%%%%%%%%%%%%%%%%%%%%%%
\begin{figure}[h]
 \epsfysize=8cm 

\centerline{\epsfbox{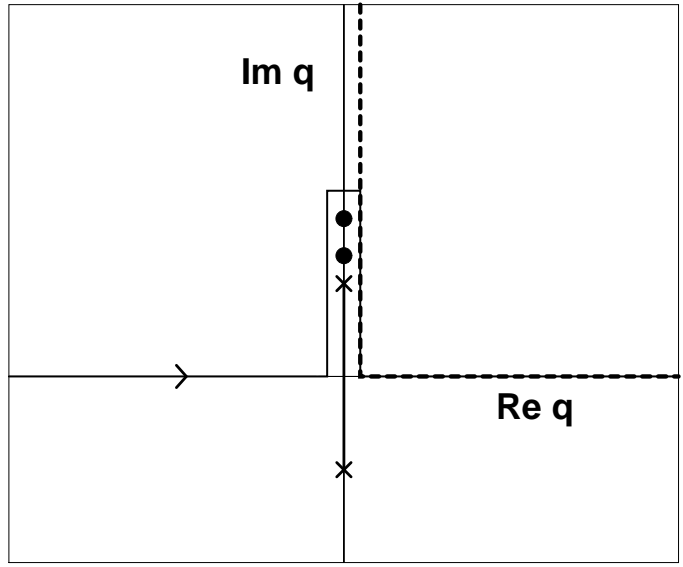}}
\caption{\small Integration contours $C$ (solid line) and $L$ (dashed 
line) in the $q$-plane.
The crosses represent branch points 
joined by a branch cut. The circles on the imaginary axis are bound states.}
\end{figure}
%%%%%%%%%%%%%%%%%%%%%%%%

Finally, we set the zero of energy at the upper level
of the potential profile using the
change of variable $E'=E-V_0=q^2/2m$, and
define the wave that satisfies the Schr\"odinger equation
for this new level in terms of the wave function written for the
lower level,
$\psi_{step}'(x,t)=e^{iV_0 t/\hbar}\psi_{step}(x,t)$, 
\beqa
\psi'_{step}(x,t)&=&\frac{1}{h^{1/2}}\int_L dq\, e^{iqx/\hbar}e^{-iE't/\hbar}
\frac{q}{p}T^l(p)\tilde\psi'_{step}(p)\nonumber
\\
&=&\frac{1}{h^{1/2}}
\intf dE'
e^{iqx/\hbar}e^{-iE't/\hbar}\frac{m}{p}T^l(p)\tilde\psi'_{step}(p)\,.
\label{bE}
\eeqa
We have thus expressed $\psi'_{step}$ in the form of the source wave
function $\psi_s$. The same arguments  used to
find the equality of  $\psi_f$ and $\psi_s$ in $D_+$
may now be applied to identify $\psi_s$ and $\psi_{step}$.  In particular,
\beq\label{rela}
\chi_s(E')=\frac{m}{p}T^l(p)\tilde\psi'_{step}(p)\,,
\eeq
where $p=\left[2m\left(E'+V_{0}\right)\right]^{1/2}$,
when
\beq
\psi_s(x=0,t)=\psi_{step}(x=0,t)\Theta(t).
\eeq
We have performed a numerical test with the simplest step potential,
Eq.~(\ref{stepp}). In this case,   
\beq\label{Tstep}
T^l(p)=\frac{2p}{p+q}.
\eeq
The state is initially localized
in the lower level of the step and sent towards the step.
It is the same state used for Figures 1 and 2, 
but boosted with a non zero momentum below the
step threshold.
Figures 4 and 5 show the real and imaginary parts of $\chi_s(E')$
calculated by Fast Fourier Transform of the signal $\psi'_{step}(x=0,t)\Theta(t)$,
which has been obtained numerically by solving the time dependent
Schr\"odinger
equation with the Split Operator Method.
Note the clear dominance of
negative energy components. The curves shown are indistinguishable from
the real and imaginary parts of the
the right hand side of  Eq.~(\ref{rela}).

%%%%%%%%%%%%%%%%%%%%%
\begin{figure}[h]
\epsfysize=8cm 
\centerline{\epsfbox{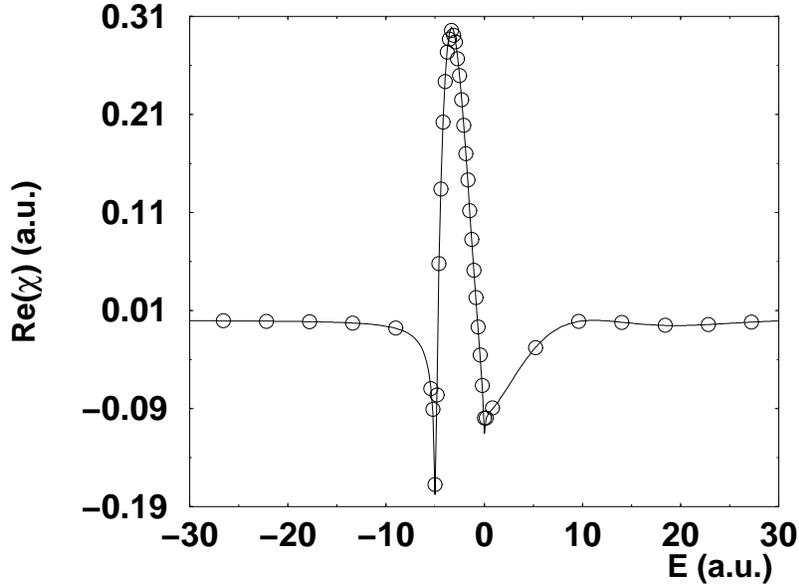}}
\caption{\small Real parts of $\chi_s(E')$ (circles) and 
$\frac{m}{p}T^l(p)\tilde\psi'_{step}(p)$ (solid line) for the step 
potential ($V_0=5$). The initial state is the same one used in 
Figures 1 and 2, but shifted in momentum so that $\la p\ra=1$.}
\label{f4}
\end{figure}
%%%%%%%%%%%%%%%%%%%%%
\begin{figure}[h]
\epsfysize=8cm 
\centerline{\epsfbox{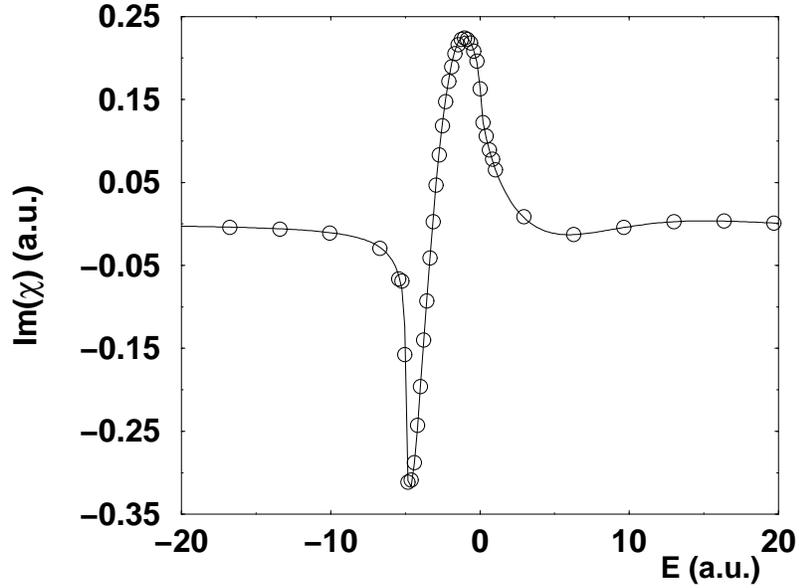}}
\caption{\small Imaginary parts of $\chi_s(E')$ (circles) and 
$\frac{m}{p}T^l(p)\tilde\psi'_{step}(p)$ (solid line) for the step 
potential ($V_0=5$). Initial state as in Figure \ref{f4}.} 
\end{figure}
%%%%%%%%%%%%%%%%%%%%%%

All the connections made in this section between the momentum amplitude
$\tilde\psi_f(p)$ and the
energy Fourier transform of the source signal at $x=0$ have required 
a wave packet strictly localized between $a$ and $b$, in order to
perform
contour deformations. It is intuitively clear,
however that relaxing this 
condition
cannot lead to a completely different behavior in the relevant energy region.
To test this claim, we have
also calculated  the real and imaginary parts of $\chi_s(E')$
by Fast Fourier Transform of $\psi'_{step}(x=0,t)\Theta(t)$,
which is obtained by solving the time dependent Schr\"odinger
equation for an initial Gaussian state with negligible overlap
with the positive-$x$ region; this is compared with the right hand side of 
Eq.~(\ref{rela}) in Figures 6 and 7. Clearly, the right hand side of Eq.~(\ref{rela}),
obtained directly from the initial state, reproduces the Fourier transform
in the important
energy region, but for very negative values of $E'$ they do not coincide. 

%%%%%%%%%%%%%%%%%%%%%%
\begin{figure}[h]
\epsfysize=8cm 
\centerline{\epsfbox{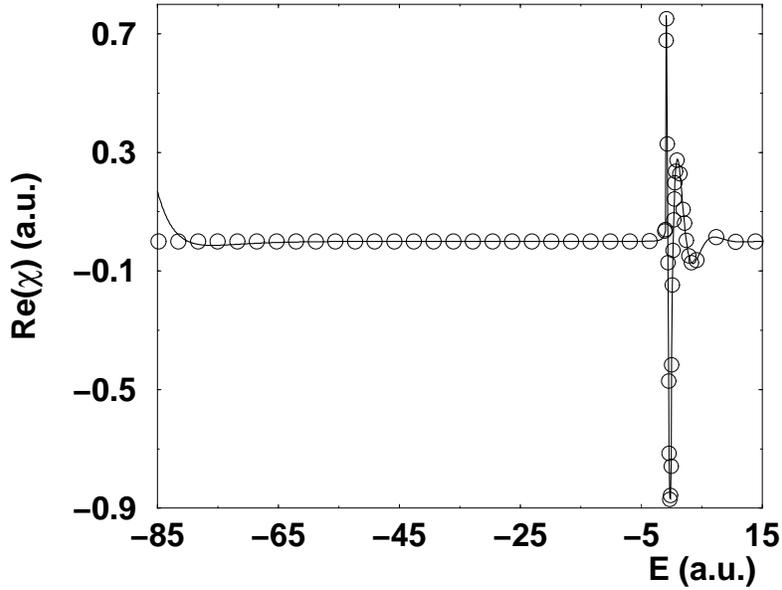}}
\caption{\small Real parts of $\chi_s(E')$ (circles) and 
$\frac{m}{p}T^l(p)\tilde\psi'_{step}(p)$ (solid line) for the step 
potential ($V_0=1$). The initial state is a minimum uncertainty-product Gaussian
with $\la x\ra=-3$, $\la p\ra=1$, and $\Delta_x=0.5$.}
\label{f6}
\end{figure}
%%%%%%%%%%%%%%%%%%%%%%%
\begin{figure}[h]
\epsfysize=8cm  
\centerline{\epsfbox{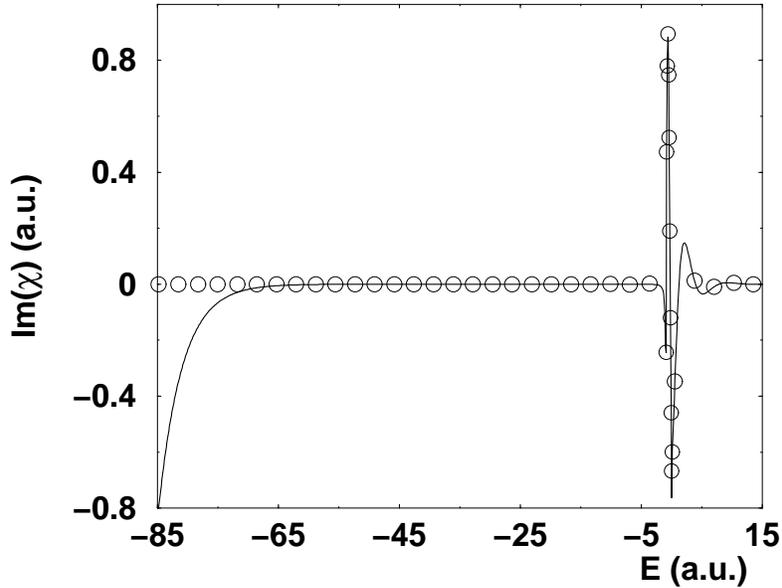}}
\caption{\small Imaginary parts of $\chi_s(E')$ (circles) and 
$\frac{m}{p}T^l(p)\tilde\psi'_{step}(p)$ (solid line) for the step 
potential ($V_0=1$). Initial state as in Figure \ref{f6}.}
\end{figure}
%%%%%%%%%%%%%%%%%%%%%%%

\section{Application to the arrival time}
Allcock obtained Eq.~(\ref{source}) by assuming that Eq.~(\ref{3.1})
was satisfied, and
that $V(x)=0$ for $x>0$, without any
further specification of the potential after the
initial state is released \cite{Allcock:1969a}.
But each experimental setup will involve a particular potential
profile, and we have seen that the interpretation of the source solutions
in terms
of an initial value problem depends on the particular potential
that determines the dynamics.
We shall in particular choose the simplest case of free motion
to analyze Allcock's objection
to the possibility of defining an ideal (apparatus independent) arrival-time
concept in quantum mechanics based on Eq.~(\ref{source}). This section is
an elaboration of the arguments given in \cite{gonrick:1999}.

Allcock's motivation for studying quantum mechanics with sources was the
fact that the expression for the wave function, Eq.~(\ref{source}),
involves an integral over energy from
$-\infty$ to $\infty$. This suggested a possible way out to
Pauli's argument against the existence of self-adjoint time operators
conjugate to a semi-bounded Hamiltonian \cite{Note}.

The core of Allcock's objection rests on the identification of
the total arrival
probability at a point
$X>0$, $P(\infty)$, with the total amount of norm
to the right of $X$ as $t\to\infty$,
\beq\label{4.4}
P(\infty)\equiv\lim_{t\to\infty}\int_X^\infty dx\,|\psi_s(x,t)|^2=
\into dE\,|\la E|\psi_s\ra|^2.
\eeq
Here
\beq
\la E|\psi_s\ra\equiv\chi_s(E)(2E/m)^{1/4},
\eeq
where, as before, the branch cut for the fourth root is set at the
negative imaginary axis.
For free motion the second equality in  Eq.~(\ref{4.4}) is to be
expected since
$\la E|\psi_s\ra(2E/m)^{1/4}$, with $E>0$, is nothing but the
amplitude for positive momentum $p$, so that we recover
from  Eq.~(\ref{4.4}) the known result
\beq
\lim_{t\to\infty}\int_X^\infty dx\,|\psi_s(x,t)|^2
=\into dp\, |\tilde\psi_f(p)|^2.
\eeq
Using two other independent routes Allcock arrives at
integral expressions for a hypothetical total arrival time probability
in contradiction
with  Eq.~(\ref{4.4})
because they have contributions from $E<0$;
the simplest one is $\intf dE |\la E|\psi_s\ra|^2$. He therefore
concludes ``{\it unequivocally that an ideal concept of arrival time
cannot be established for the problem with sources $(-\infty<E<\infty)$}''.
In the light of the identification between $\psi_s(x,t)$ and $\psi_f(x,t)$
in $D_+$ made in section II, the flaw in Allcock's argument is that he is
overlooking the contribution of negative momenta. According to
classical prejudice one would not expect that negative momenta play any role
for $t>0$ at $X>0$, if the initial state is confined within the left half-line.
But quantum mechanics works otherwise, as
Figures 1 and 2 clearly show.
Since the negative energy part of $\psi_s(x,t)$
is equal to $\psi_{f,-}(x,t)$,
we may equivalently think of the contribution of the negative momenta
to a time-of-arrival probability in terms of evanescent waves.
The negative energy components penetrate the positive coordinate region
for some time before
finally abandoning it. But Allcock's definition of $P(\infty)$ disregards
this contribution entirely. Note that our criticism to the validity of
$P(\infty)$ is rather independent
on the theory chosen for defining an arrival time distribution.
However, we cannot fail to point out
that the equality between $\psi_s$ and $\psi_f$ in the space-time domain
$D_+$ allows immediately to associate Kijowski's distribution $\Pi_K$
\cite{Kijowski:1974jx,gonrick:1999} to the source problem for free motion.
This arrival time distribution
has been extensively discussed elsewhere
\cite{Muga:1998gw,Baute:1999hb,Egusquiza:2000a}.
Similarly, for a general potential profile
the generalization of $\Pi_K$ proposed in \cite{Baute:1999hb} is applicable.

\section{Discussion}

The physical meaning of source boundary conditions used for the
one dimensional, time dependent, single particle Schr\"odinger
equation has been clarified by showing the equivalence of the
corresponding solution $\psi_s$, see Eq.~(\ref{source}),
with standard initial value problems for positive times and positions.
We have first shown the connection between Eq.~(\ref{source}) and
the ordinary initial-value-problem integral expression for a free motion
wave packet,
and then have extended this result for other potential profiles,
in particular for the step barriers, which allow to inject waves
dominated by evanescent energies into the right half-line.

Each of the potential profiles may correspond to a different experimental
arrangement.
Depending on the potential the origin of the evanescent
waves in the source solution may be traced back to either
negative momentum
components (which are present in all cases, and become the only
source of negative energies for free-motion),
or ``true'' evanescent components, namely energy components which are
already negative in the associated initial value problem,
relative to the region of positive $x$. 
This occurs for example
in the step potentials,
where positive momenta with respect to the lower level become imaginary
momenta in the upper level region.
In addition to scattering contributions to the negative energies, we have
also allowed for the possibility of bound state contributions, that may become
important if the initial state overlaps with the normalized
eigenstates of the Hamiltonian.

Apart from elucidating the physical content of
previous works that have made use of source boundary conditions,
our results have also served to  analyze  critically
Allcock's negative statements about the possibility to define
an ideal probability distribution of time of arrival for
quantum waves with sources. It is argued that Allcock's postulated
expression for a total arrival probability is flawed since it ignores the
quantum contribution of negative momenta at positive positions and times.

\acknowledgments{We are grateful to Jos\'e Pascual Palao 
for providing us with numerical tools. 
This work has been supported
by Ministerio de Educaci\'on
y Cultura (Grants
PB97-1482 and AEN99-0315), The
University of the Basque Country (Grant UPV 063.310-EB187/98), 
and the Basque Government (Grant PI-1999-28).
A. D. Baute acknowledges an FPI fellowship by Ministerio de
Educaci\'on y Cultura.}

\appendix
\section{Solvable example}
We shall work out the time dependence of a state which is at $t=0$
the ground state of an infinite well of width $D$ between $a$ and $b=a+D$,
and evolves freely from that time on.
This was the only example of source mentioned by Allcock
\cite{Allcock:1969a}: 
\beq\label{psiiv}
\psi_{iv}(x,t=0)=\left(\frac{2}{D}\right)^{1/2}\sin[(x-a)k_w]{\cal H}(a,b),
\eeq
where $k_w=\pi/D$, and
\beq
{\cal H}(a,b)=\cases{
1&$\,\,\, a<x<b$\cr
0&$\,\,\, {\rm otherwise}$\cr}.
\eeq
Eq. (\ref{psiiv}) may also be written as a combination of four plane waves
truncated at $a$ and $b$, two of them
with momentum $p_w=k_w \hbar$, and the other two with momentum $-p_w$. 
The time evolution of a truncated plane wave was first solved by
Moshinsky \cite{Moshinsky:1952}. For an alternative derivation see
\cite{Brouard:1996}.
The resulting wave is given by
\beq
\psi_{iv}(x,t)=\frac{1}{4i}\left(\frac{2}{D}\right)^{1/2}
\left\{e^{ik_w D}e^{im(x-b)^2/2t\hbar}[w(-u_{b+})-w(-u_{b-})]
+e^{im(x-a)^2/2t\hbar}[w(-u_{a-})-w(-u_{a+})]\right\},
\eeq
where
\beqa
u_{b\pm}&=&\frac{\pm p_w-m(x-b)/t}{f},
\nonumber\\
u_{a\pm}&=&\frac{\pm p_w-m(x-a)/t}{f},
\nonumber\\
f&=&(1-i)\left(\frac{m\hbar}{t}\right)^{1/2},
\nonumber\\
w(u)&=&e^{-u^2}{\rm erfc}(-iu).
\eeqa
For properties of the $w$-function see \cite{Abramowitz:1972}.
The momentum representation of  Eq.~(\ref{psiiv}) is also of interest
to illustrate the properties mentioned in section II (analyticity
and decay to zero in the upper half plane when the state is
confined between two points in coordinate space),
\beq
\tilde\psi_{iv}(p)=-\frac{(2h)^{1/2}}{4\pi D^{1/2}}
\sum_{\alpha=\pm} \alpha \frac{e^{-i(p+\alpha p_w)b/\hbar}-e^{-i(p+\alpha
p_w)a/\hbar}}
{p+\alpha p_w}
e^{i\alpha p_w a/\hbar}.  
\eeq

\end{document}